\documentclass[12pt]{article}

\catcode`\@=11

\global\arraycolsep=2pt
\usepackage{amsbsy,amssymb,latexsym,amsfonts,amsmath}
\usepackage{graphicx,color}
\usepackage{tikz}
\usetikzlibrary{arrows.meta,decorations.pathmorphing,decorations.markings,decorations.pathreplacing,calc}
\tikzset{
  axisline/.style={-{Stealth[length=2mm]},gray!70,thin},
  contour/.style={blue!70!black,very thick,
    decoration={markings,
      mark=at position 0.28 with {\arrow{Stealth[length=2.2mm]}},
      mark=at position 0.75 with {\arrow{Stealth[length=2.2mm]}}},
    postaction={decorate}},
  sdpath/.style={red!75!black,thick,dashed},
  cutline/.style={brown!70!black,line width=1pt,
    decorate,decoration={snake,amplitude=0.7mm,segment length=2.2mm}},
  saddle/.style={circle,fill=black,inner sep=1.6pt},
  ghost/.style={circle,draw=black!45,fill=white,inner sep=1.5pt,line width=0.8pt},
}
\usepackage[numbers]{natbib}

\newcommand{\gc}{\gamma_c}
\newcommand{\Det}{\operatorname{Det}}
\providecommand{\Tr}{\operatorname{Tr}}
\newcommand{\dd}{\mathrm{d}}
\newcommand{\ee}{\mathrm{e}}

\newcommand{\sech}{\operatorname{sech}}

\usepackage{hyperref}
\hypersetup{
    colorlinks=true,
    linkcolor=blue,
    citecolor=blue,
    urlcolor=blue
}

\begin{document}

\begin{flushright}
YITP-26-111
\end{flushright}

\vspace*{0.7cm}

\begin{center}
{ \Large Semiclassical Liouville Theory on the Real Projective Plane:\\[2mm] A Complex Interpretation of the Bootstrap}
\vspace*{1.5cm}\\
{Yu Nakayama}
\end{center}
\vspace*{1.0cm}
\begin{center}

Yukawa Institute for Theoretical Physics,
Kyoto University, Kitashirakawa Oiwakecho, Sakyo-ku, Kyoto 606-8502, Japan

\vspace{3.8cm}
\end{center}

\begin{abstract}
The exact one-point function of Liouville theory on the real projective plane was derived long ago from the bootstrap, yet it has never been verified against a semiclassical path-integral computation. The check is less straightforward than one might expect. On the real projective plane every constant-curvature metric is positively curved, whereas classical Liouville theory produces metrics of constant negative curvature, so no real classical solution exists. Instead, as we show, the semiclassical path integral receives contributions from infinitely many complex saddles with a negative-definite metric. Once the integration contour is chosen so that the path integral converges, these saddles reproduce the exact one-point function. Complex saddles arise in the same way in timelike Liouville theory and in two-dimensional de~Sitter gravity, and the example treated here offers a rare opportunity to test their use against a known exact answer.
\end{abstract}

\setcounter{page}{0}
\thispagestyle{empty}

\newpage

\section{Introduction}

Quantum Liouville theory and its relatives have played a distinctive role in theoretical physics. They appear in two-dimensional quantum gravity, in non-critical string theory, in the $\mathrm{AdS}_3/\mathrm{CFT}_2$ correspondence, in Calabi--Yau compactifications, and beyond (see~\cite{Nakayama} for a review). The hallmark of Liouville theory is that it is non-rational and yet exactly solvable. Non-rational conformal field theories are believed to have universal conformal data at large conformal dimension. These data are given by the structure constants of Liouville theory~\cite{CMMT}, and they are expected to describe holographic conformal field theories in an averaged sense~\cite{CCHM}. The corresponding data for unoriented surfaces are encoded in the one-point function on the real projective plane $\mathbb{RP}^2$, or equivalently in the crosscap state~\cite{Tsiares}. This one-point function is the subject of the present paper.

The boundary states of Liouville theory associated with the ZZ~\cite{ZZpseudosphere} and FZZT~\cite{FZZ,TeschnerBoundary} boundary conditions have been studied extensively. Their exact one-point functions were obtained from the conformal bootstrap and checked against semiclassical computations~\cite{ZZpseudosphere,MenottiTonniI,MenottiTonniII}. Such a check is important because it confirms that the bootstrap solution is the one given by the path integral. See also~\cite{AbdallaChandraWang} for a semiclassical analysis of line defects in Liouville theory realized as localized cosmological constants.

Quantum Liouville theory on $\mathbb{RP}^2$ was first studied by Hikida~\cite{Hikida}, who derived the exact one-point function of the bulk vertex operator, or equivalently the crosscap wave function, from the bootstrap. The sign of this one-point function was later corrected by Bergman and Hirano~\cite{BergmanHirano} (see also~\cite{Nakayama:2003ep,Nakayama}). To the author's knowledge, however, this one-point function has not been derived from a semiclassical path-integral computation. A heuristic evaluation that retains only the Liouville zero mode, the constant part of the field, appears in~\cite{BergmanHirano}, but the saddle itself and the fluctuations around it have not been analyzed. In this paper we carry out this computation.

At first sight this looks like a routine exercise, but it is not, for two reasons. The first is the absence of a real saddle. As Liouville himself knew, the classical Liouville equation produces metrics of constant negative curvature. The real projective plane, by contrast, has positive Euler characteristic. By the Gauss--Bonnet theorem any constant-curvature metric on it must be positively curved. This remains true in the presence of the conical singularity created by the vertex operator, because the Seiberg bound~\cite{Seiberg} implies that its deficit angle is smaller than $2\pi$. Hence there is no real classical solution, and the semiclassical limit must be controlled by complex saddles. The second is the sign of the one-point function. In the probabilistic construction of Liouville theory on the sphere~\cite{DKRV,KRV} (see~\cite{ChatterjeeWitten} for a review), $\ee^{2\alpha\phi}$ is a positive random variable for real $\alpha$. If a construction of the same kind existed on $\mathbb{RP}^2$, the one-point function $\langle \ee^{2\alpha\phi}\rangle$ would also be positive. However, the exact $\langle \ee^{2\alpha\phi}\rangle$ oscillates rapidly in $\alpha$ as $b\to0$ and takes negative values. Therefore no such construction is available, and the one-point function must be defined by analytic continuation~\cite{HMW}.

We show that, once complex saddles and the analytic continuation are taken into account, the semiclassical path integral reproduces the full $\alpha$-dependence of the exact one-point function through one-loop order. The saddles have a negative-definite metric, so their negative curvature is consistent with the Gauss--Bonnet theorem. The oscillations come from infinitely many such saddles, which have the same metric and differ only in the branch of the multivalued Liouville field. Their contributions combine into a single contour integral over the zero mode. Our semiclassical result agrees with the exact one-point function up to an overall constant, which we do not compute, and a finite renormalization of the cosmological constant, which we fix by matching the normalization used in the exact formula.

This paper contains no new exact results. Its purpose is to show, in a case where the exact answer is known, how the analytic continuation of Liouville theory~\cite{HMW} should be implemented in the path integral. This analytic continuation underlies much recent work, in particular on timelike Liouville theory. The timelike three-point function, the $c\le1$ analog of the DOZZ formula, was obtained and studied in~\cite{StromingerTakayanagi,McElgin,GiribetTimelike,RibaultSantachiara,BautistaErbinKudrna,KapecMahajan}. On the mathematical side, rigorous probabilistic constructions and integrability results have been obtained for timelike (or imaginary) Liouville theory~\cite{AngCaiSunWu,GKRimaginary,Chatterjee,UsciatiGKS,Chatterjee2}. Coupling two Liouville theories with complex-conjugate central charges defines the complex Liouville string~\cite{ComplexLiouvilleString,CLSworldsheet,CLSmatrix,CLSboundaries,CLSgravity}, a solvable model of two-dimensional quantum gravity that is dual to a two-matrix integral and equivalent to sine-dilaton gravity. A closely related model built from spacelike and timelike Liouville theories is the Virasoro minimal string~\cite{VirasoroMinimalString}. The same analytic continuation is required when Liouville theory is used to describe lower-dimensional holography, de~Sitter quantum gravity, and two-dimensional cosmology~\cite{BautistaDabholkarErbin,AnninosTwoSphere,VerlindeZhang,AnninosBaraccoMuhlmann,AllamehShaghoulian,LiouvilleCosmology,GiribetSivilotti,ComplexSaddlesdS3,ComplexSaddlesCS3,Semiclassical3dSaddlesLetter,Semiclassical3dSaddles,HondaShinmyo}.

The rest of the paper is organized as follows. In section~\ref{sec:setup} we review the bootstrap formula and extract the semiclassical limit that our computation must reproduce. In section~\ref{sec:saddle} we evaluate the classical action of the complex saddles, in section~\ref{sec:oneloop} we compute the fluctuation determinant, and in section~\ref{sec:zeromode} we perform the zero-mode integral. We discuss the results in section~\ref{sec:summary}. In appendix~\ref{app:action} we evaluate the four terms of the classical action, in appendix~\ref{app:gy} we state the Gelfand--Yaglom theorem in the form needed for the crosscap boundary conditions, and in appendix~\ref{app:renorm} we provide an independent evaluation of the fluctuation determinant.

\section{Setup and semiclassical limit}
\label{sec:setup}

\subsection{Action and semiclassical variables}\label{sec:variables}

We consider the one-point function of a bulk vertex operator on the real projective plane, formally defined by the path integral $\langle \ee^{2\alpha \phi(0)} \rangle = \int \mathcal{D} \phi\, \ee^{2\alpha \phi(0)}\, \ee^{-S[\phi]}$ with the insertion at the origin. The Liouville action is
\begin{equation}\label{eq:Sphi}
S[\phi]=\frac1\pi\int\dd^2z\,\partial\phi\,\bar\partial\phi
+\mu\int\dd^2z\,\sqrt{\hat g}\,\ee^{2b\phi}
+\frac{Q}{4\pi}\int\sqrt{\hat g}\,\hat R\,\phi .
\end{equation}
Here $\mu>0$ is the cosmological constant, $\hat g$ is a fiducial metric on $\mathbb{RP}^2$, and $\hat R$ is its curvature.\footnote{Our conventions follow those of~\cite{Nakayama}. In particular $\dd^2z=\dd x\,\dd y$ and $\delta^2(z)=\delta(x)\delta(y)$, which differ from those of Polchinski's textbook~\cite{PolchinskiBook}. The Liouville literature traditionally sets $\alpha'=1$, which fixes the normalization of the kinetic term.} Since we are interested in the limit $b\to0$, we rescale the field to $\varphi\equiv2b\phi$ and hold $\Lambda\equiv\pi\mu b^2$ fixed, so that the action takes the form
\begin{align}\label{eq:action}
S[\phi]=\frac1{b^2}\left(\frac1{4\pi}\int\dd^2z\,\partial\varphi\,\bar\partial\varphi
+\frac{\Lambda}{\pi}\int\dd^2z\,\sqrt{\hat g}\,\ee^\varphi\right)
+\frac{Q}{8\pi b}\int\sqrt{\hat g}\,\hat R\,\varphi,\qquad Q=b+\frac1b .
\end{align}
The terms of order $1/b^2$, including the $O(b^{-2})$ part of the background-charge term, form the classical action. The background-charge term also has an $O(1)$ part, which contributes at one loop.

The space $\mathbb{RP}^2$ is the quotient of the sphere $\mathbb{S}^2$ by the antipodal map $\sigma:z\mapsto-1/\bar z$, which in polar coordinates is $(r,\theta)\mapsto(1/r,\theta+\pi)$. This map has no fixed points, maps the unit circle to itself, and exchanges its interior and exterior, so the unit disk is a fundamental domain for $\sigma$, and $\mathbb{RP}^2$ is this disk with antipodal points of the circle $r=1$ identified. Since the fiducial metric can be any metric on $\mathbb{RP}^2$, we take the flat metric $\dd\hat s^2=|\dd z|^2=\dd r^2+r^2\dd\theta^2$ on the disk. The bulk curvature then vanishes, so the whole curvature term is concentrated on the circle $r=1$, across which the fiducial metric is not smooth. There it reduces to an integral of the geodesic curvature, which we evaluate in appendix~\ref{app:action}.

The vertex operator $V_\alpha=\ee^{2\alpha\phi}=\ee^{\alpha\varphi/b}$ has conformal dimension $\Delta=\alpha(Q-\alpha)$. In the semiclassical limit we take it to be heavy,
\begin{equation}
\alpha=\frac{\eta}{b}\quad(\eta\ \text{fixed},\ b\to0),\qquad
\gc\equiv1-2\eta,
\end{equation}
so that it acts as a source of strength $\eta/b^2$ for the classical field. The Liouville momentum $\alpha$ is restricted by the Seiberg bound $\alpha\le Q/2$~\cite{Seiberg}, which in these variables reads $\eta\le\tfrac12$, or $\gc\ge0$. We work throughout in the interior of this range, $0<\eta<\tfrac12$, i.e., $0<\gc<1$.

In the physical metric $\dd s^2=\ee^{\varphi}\,\dd\hat s^2$ the insertion point is a conical singularity, whose cone angle is determined by the classical solution of section~\ref{sec:saddle}. On the double cover $\mathbb{S}^2$ the insertion at $r=0$ is accompanied by its antipodal image at $r=\infty$, with the same Liouville momentum $\alpha$. We will use the double cover in appendix~\ref{app:renorm}, where the Green function on $\mathbb{RP}^2$ is obtained from the one on the sphere by the method of images.

\subsection{The bootstrap formula and its semiclassical limit}

The exact crosscap one-point function was obtained by Hikida~\cite{Hikida} from the modular bootstrap applied to the M\"obius-strip amplitude between the ZZ brane and the crosscap state, with the sign corrected in~\cite{BergmanHirano} (see also~\cite{Nakayama:2003ep,Nakayama}). Crossing symmetry of the bulk two-point function on $\mathbb{RP}^2$ provides an independent check~\cite{FioravantiPradisiSagnotti,PradisiSagnottiStanevWZW,PradisiSagnottiStanev,Nakayama:2016cim}. With the flat fiducial metric of section~\ref{sec:variables}, which we use in our semiclassical computation, the formula reads
\begin{align}\label{eq:U}
U(\alpha)&=\frac2b\big(\pi\mu\gamma(b^2)\big)^{\frac{Q-2\alpha}{2b}}
\Gamma(2b\alpha-b^2)\,\Gamma\!\Big(\frac{2\alpha}{b}-\frac1{b^2}-1\Big)\,f(\alpha), \\
f(\alpha)&=2\cos\!\big[\pi b(\alpha{-}\tfrac Q2)\big]\cos\!\big[\tfrac\pi b(\alpha{-}\tfrac Q2)\big],
\end{align}
where $\gamma(x)=\Gamma(x)/\Gamma(1-x)$.

For real $\eta$ and $b$ every factor in \eqref{eq:U} is real, but as $b\to0$ the function $U$ oscillates rapidly in $\alpha$ and changes sign. A single real saddle cannot produce this behavior. We will see that the oscillation is entirely contained in a factor $1/\cos\pi\nu$, with $\nu$ defined in \eqref{eq:nuneg_exp}, while the remaining factor has an ordinary semiclassical expansion. We therefore write the semiclassical form of \eqref{eq:U} as
\begin{equation}\label{eq:nuneg_exp}
\log\big|U(\alpha)\,\cos\pi\nu\big|=\frac{-S_{\mathrm{cl}}}{b^2}+W_0+W_1+O(b^2),
\qquad
\nu\equiv\frac{\gc}{2b^2},
\end{equation}
with the classical part
\begin{equation}\label{eq:nuneg_action}
-S_{\mathrm{cl}}=\frac\gc2\log\Lambda-\gc\log\gc+\gc ,
\end{equation}
which includes the contribution of the source, and the one-loop term
\begin{equation}\label{eq:nuneg_W1}
W_1=\log\Gamma(2\eta)+\log\sin\pi\eta-\tfrac32\log\gc-\gc\gamma_E ,
\end{equation}
where $\gamma_E$ is Euler's constant. Both $S_{\mathrm{cl}}$ and $W_1$ are real for $0<\eta<\tfrac12$. We have collected the remaining $\eta$-independent terms in the normalization $W_0$.

To extract this structure from \eqref{eq:U}, we isolate the oscillating factor with the $\Gamma$-function identity
\begin{equation}\label{eq:Gf}
\Gamma\!\Big(-\frac\gc{b^2}-1\Big)f(\alpha)
=-\frac{\pi\,\cos\!\big[\tfrac\pi2(\gc+b^2)\big]}{\cos\pi\nu\,\Gamma\!\big(\tfrac\gc{b^2}+2\big)},
\end{equation}
which follows from $f(\alpha)=-2\cos[\tfrac\pi2(\gc+b^2)]\sin\pi\nu$ and the reflection formula for the $\Gamma$-function. The remaining factor $\Gamma(\gc/b^2+2)$ is smooth, and its Stirling expansion contributes to both the classical action \eqref{eq:nuneg_action} and the one-loop term \eqref{eq:nuneg_W1}.

The factor $1/\cos\pi\nu$ can be interpreted as a sum over infinitely many complex saddles. To see this, we give $\nu$ a small positive imaginary part, which amounts to giving $b^2$ a small negative imaginary part. Then $1/\cos\pi\nu$ admits the convergent expansion
\begin{equation}\label{eq:cosexp}
\frac1{\cos\pi\nu}=2\sum_{n\ge0}(-1)^n\,\ee^{i(2n+1)\pi\nu}\qquad(\operatorname{Im}\nu>0).
\end{equation}
The opposite choice of sign gives the complex-conjugate series, and both represent the same function. Comparing the $n$-th term with the semiclassical form $\log U=-S_{\mathrm{cl}}/b^2+\cdots$, we see that it has the form of a saddle-point contribution with action
\begin{equation}\label{eq:treetarget}
\frac{-S_{\mathrm{cl}}^{(n)}}{b^2}
=\underbrace{\frac1{b^2}\Big(\frac\gc2\log\Lambda-\gc\log\gc+\gc\Big)}_{n\text{-independent}}
\ +\ (2n+1)\,i\pi\nu .
\end{equation}
For real $\nu$, the real part of \eqref{eq:treetarget} is the same for every $n$ and coincides with the classical term $-S_{\mathrm{cl}}/b^2$ of \eqref{eq:nuneg_exp}, so all terms in \eqref{eq:cosexp} have the same absolute value and differ only in the phase factor $\ee^{(2n+1)i\pi\nu}$ and the sign $(-1)^n$.

The one-loop term \eqref{eq:nuneg_W1} is real and common to all saddles, so the one-point function factorizes into a smooth real factor and the sum over phases \eqref{eq:cosexp}. In the rest of the paper we derive both factors from the path integral.

\section{Complex semiclassical saddles}
\label{sec:saddle}

\subsection{Saddle solution}
\label{sec:saddlesol}

Varying the classical action of \eqref{eq:action}, supplemented by the source $\tfrac{\eta}{b^2}\varphi(0)$ from the heavy insertion, we obtain the Liouville equation with a delta-function source,
\begin{equation}\label{eq:eom}
\partial\bar\partial\varphi=2\Lambda\,\ee^{\varphi}-2\pi\eta\,\delta^2(z).
\end{equation}
Near the origin the delta-function source imposes $\varphi\sim-4\eta\log r$, which corresponds to a conical singularity of the physical metric at the insertion. Since the problem is rotationally symmetric about the insertion, we look for a solution $\varphi=\varphi(r)$. Away from $r=0$, equation \eqref{eq:eom} then reduces to the radial equation
\begin{equation}\label{eq:eom-radial}
\varphi''+\frac{\varphi'}{r}=8\Lambda\,\ee^{\varphi}.
\end{equation}
Its general solution is
\begin{equation}\label{eq:gen-sol}
\ee^{-\varphi/2}=c_+\,r^{1+\beta}+c_-\,r^{1-\beta},\qquad c_+c_-\,\beta^2=-\Lambda,
\end{equation}
which contains two free constants, since the product $c_+c_-$ is fixed by the constraint. We take them to be $\beta$ and $c_+/c_-$ and determine them from the boundary conditions at $r=0$ and $r=1$. At $r=0$ the term $r^{1-\beta}$ dominates for $\beta>0$ and gives $\varphi\sim-2(1-\beta)\log r$, which reproduces the behavior $\varphi\sim-4\eta\log r$ imposed by the source precisely when
\begin{equation}\label{eq:beta-fix}
\beta=\gc .
\end{equation}
The constraint then reads $c_+c_-=-\Lambda/\gc^2$, and one constant remains to be fixed at $r=1$.

At $r=1$ the solution must respect the crosscap identification, which means that the physical metric $\dd s^2=\ee^{\varphi}|\dd z|^2$ must be invariant under the antipodal map $\sigma:z\mapsto-1/\bar z$, so that it descends from the sphere to $\mathbb{RP}^2$. Since $|\dd(\sigma z)|^2=|z|^{-4}|\dd z|^2$, invariance requires
\begin{equation}\label{eq:equiv}
\varphi(1/r)=\varphi(r)+4\log r .
\end{equation}
Imposed on \eqref{eq:gen-sol} with $\beta=\gc$, this condition gives $c_+=c_-$, and combined with the constraint it yields
\begin{equation}\label{eq:csq}
c_+^2=-\frac{\Lambda}{\gc^2}<0 .
\end{equation}
This equation has no real solution, as anticipated in the introduction, so the saddle is necessarily complex.

\subsection{The saddle field and its branches}

Choosing the root $c_+=c_-\equiv c=i\sqrt\Lambda/\gc$ of \eqref{eq:csq} and taking the logarithm of \eqref{eq:gen-sol}, we obtain the saddle field
\begin{equation}\label{eq:phi-star}
\varphi_*(r) =-2\log\!\Big[\frac{i\sqrt\Lambda}{\gc}\big(r^{1+\gc}+r^{1-\gc}\big)\Big],
\end{equation}
which we write as $\varphi_*=-2\log c-2\log g(r)$ with $g(r)\equiv r^{1+\gc}+r^{1-\gc}=2r\cosh(\gc\log r)$.

The Weyl factor of the physical metric $\ee^{\varphi_*}|\dd z|^2$ is
\begin{equation}\label{eq:metric}
\ee^{\varphi_*}=-\frac{\gc^2}{\Lambda}\,\big(r^{1+\gc}+r^{1-\gc}\big)^{-2},
\end{equation}
in which the constant $c$ appears only through $c^2=-\Lambda/\gc^2$. The Weyl factor is real and single-valued, but negative. Since reversing the sign of the metric reverses the sign of its curvature, the saddle is consistent with the Gauss--Bonnet theorem. Up to this overall sign, the physical metric is that of the antipodal quotient of a spindle. A spindle is a sphere with a metric of constant positive curvature and two conical singularities, located here at $r=0$ and $r=\infty$. The overall sign does not affect the conical singularity at the insertion, where $\ee^{\varphi_*}\simeq-\tfrac{\gc^2}{\Lambda}\,r^{-2(1-\gc)}$ corresponds to a cone angle of $2\pi\gc$, i.e., a deficit angle of $4\pi\eta$.

The field $\varphi_*$, on the other hand, is multivalued. Since only $c^2$ is fixed by \eqref{eq:csq} and $c$ is purely imaginary, the choice of the sign of $c$ and of the branch of $\log c$ gives a discrete family of saddles,
\begin{equation}\label{eq:phi-tower}
\varphi_*^{(n)}(r)=\varphi_*^{(0)}(r)-2\pi i\,n,\qquad n\in\mathbb{Z}.
\end{equation}
We take $\varphi_*^{(0)}$ to be the branch with $c=i\sqrt\Lambda/\gc$, whose additive constant $-2\log c$ has imaginary part $-\pi$. All members of the family share the Weyl factor \eqref{eq:metric} and differ only by the additive imaginary constant, but as field configurations they are distinct saddles of the complex path integral. The additive constant affects the classical action only through the terms linear in $\varphi_*$, namely the source term $\eta\,\varphi_*(0)$ and the curvature term $\tfrac1{8\pi}\int\sqrt{\hat g}\,\hat R\,\varphi_*$. The cosmological-constant and kinetic terms involve only $\ee^{\varphi_*}$ and $\varphi_*'$ and are the same for every $n$. We keep the whole family and compute the classical action of each member.

\subsection{Classical action}

In the semiclassical limit the one-point function is a sum over the saddles \eqref{eq:phi-tower},
\begin{equation}\label{eq:logU-Scl}
U=\sum_{n}w_n\,\exp\!\Big(\frac{-S_{\mathrm{cl}}^{(n)}}{b^2}+W_1+\cdots\Big),
\qquad
-S_{\mathrm{cl}}^{(n)}=\eta\,\varphi_*^{(n)}(0)-\mathcal{S}[\varphi_*^{(n)}],
\end{equation}
where $\eta\,\varphi_*^{(n)}(0)$ is the contribution of the insertion $V_\alpha=\ee^{\eta\varphi/b^2}$ and $\mathcal{S}$ is the coefficient of $1/b^2$ in the action \eqref{eq:action},
\begin{equation}\label{eq:reduced}
\mathcal{S}[\varphi]=\frac1{4\pi}\int\dd^2z\,\partial\varphi\,\bar\partial\varphi
+\frac{\Lambda}{\pi}\int\dd^2z\,\ee^{\varphi}
+\frac1{8\pi}\int\sqrt{\hat g}\,\hat R\,\varphi . \end{equation} Its last term is the $O(b^{-2})$ part of the background-charge coupling. We compute the one-loop factors in section~\ref{sec:oneloop} and collect them into $W_1$ in section~\ref{sec:zeromode}, where we also determine the weights $w_n$.

Both the source term and the kinetic term diverge at the insertion, so we regulate the insertion at $r=\epsilon$ and write $\varphi_*(0)$ for the finite part of $\varphi_*(\epsilon)$. We evaluate the four contributions to $-S_{\mathrm{cl}}^{(n)}$ in appendix~\ref{app:action} and find
\begin{equation}\label{eq:terms}
\begin{aligned}
\text{source:}\quad &\eta\,\varphi_*^{(n)}(\epsilon)=-2\eta\log c-4\eta^2\log\epsilon ,\\
\text{kinetic:}\quad &\frac1{4\pi}\int\dd^2z\,\partial\varphi_*\bar\partial\varphi_*
=-2\eta^2\log\epsilon+\log2-\frac\gc2 ,\\
\text{cosmological:}\quad &\frac{\Lambda}{\pi}\int\dd^2z\,\ee^{\varphi_*}=-\frac\gc2 ,\\
\text{curvature:}\quad &\frac1{8\pi}\int\sqrt{\hat g}\,\hat R\,\varphi_*=\frac12\,\varphi_*^{(n)}(1)=-\log c-\log2 .
\end{aligned}
\end{equation}
The divergences of the source and kinetic terms combine in $-S_{\mathrm{cl}}^{(n)}=\eta\varphi_*(\epsilon)-\mathcal{S}$ into a residual $-2\eta^2\log\epsilon$. This residual is canceled by the normal-ordering counterterm $2\eta^2\log\epsilon$ of the vertex operator $V_\alpha$.\footnote{Since the kinetic term in \eqref{eq:Sphi} gives $\langle\phi(z)\phi(w)\rangle\simeq-\log|z-w|$ at short distances, normal ordering with a short-distance cutoff $\epsilon$ multiplies $\ee^{2\alpha\phi}$ by $\epsilon^{2\alpha^2}=\exp(2\eta^2\log\epsilon/b^2)$.} As in the semiclassical computations of~\cite{MenottiTonniI,MenottiTonniII}, we define the normal ordering with respect to the flat fiducial metric. Collecting the finite parts, we are left with
\begin{equation}\label{eq:Scl-assemble}
-S_{\mathrm{cl}}^{(n)}=\gc\log c+\gc .
\end{equation}
Substituting $\log c=\tfrac12\log\Lambda-\log\gc+\tfrac{i\pi}2(2n+1)$, where the last term distinguishes the members of the family, we find
\begin{equation}\label{eq:Scl-tower}
\frac{-S_{\mathrm{cl}}^{(n)}}{b^2}
=\underbrace{\frac1{b^2}\Big(\frac\gc2\log\Lambda-\gc\log\gc+\gc\Big)}_{\text{real, common to all }n}
\;+\;(2n+1)\,i\pi\nu ,
\qquad \nu=\frac\gc{2b^2}.
\end{equation}

This agrees with the action \eqref{eq:treetarget} obtained from the exact formula, with $n$ identified with the label of the terms in \eqref{eq:cosexp}. In particular, the imaginary parts of the classical actions reproduce the phases of these terms. The classical action alone, however, does not tell us which members of the family contribute and with what weights. The contributing saddles and their weights are determined by the zero-mode integral in section~\ref{sec:zeromode}.

\subsection{Fixed-area action}
\label{sec:fa}

We treat the zero-mode integral as an integral over the total area $A=\int_{r\le1}\ee^\varphi\,\dd^2z$, which requires the classical action at fixed area. We define it by extremizing the classical action, including the source term, subject to the area constraint,
\begin{equation}\label{eq:fa_def}
-S_{\mathrm{cl}}(A)=\operatorname*{ext}_{\int_{r\le1}\ee^{\varphi}\,\dd^2z\,=\,A}
\Big[\,\eta\,\varphi(0)-\mathcal{S}^{(0)}[\varphi]\,\Big],
\qquad
\mathcal{S}^{(0)}=\mathcal{S}-\frac{\Lambda}{\pi}\int\ee^{\varphi} ,
\end{equation}
where $\mathcal{S}$ is given in \eqref{eq:reduced}. The cosmological-constant term is omitted from $\mathcal{S}^{(0)}$ because it equals $(\Lambda/\pi)A$ on the constraint surface and does not affect the extremization. It reappears as the factor $\ee^{-\mu A}$ in the area integral \eqref{Ltranform} below.

The fixed-area saddle solves the radial equation \eqref{eq:eom-radial} with $\Lambda$ replaced by the Lagrange multiplier $\Lambda_{\rm eff}$ that enforces the constraint, so it is again of the form \eqref{eq:gen-sol} with $c_+=c_-$, now with $c^2=-\Lambda_{\rm eff}/\gc^2$. The constraint fixes the Lagrange multiplier in terms of the area,
\begin{equation}\label{eq:fa_Leff}
A=\int_{r\le1}\ee^{\varphi_*}\dd^2z=-\frac{\pi\gc}{2\Lambda_{\rm eff}}
\quad\Longrightarrow\quad
c^2=\frac{\pi}{2A\gc}.
\end{equation}
For $A>0$ the Lagrange multiplier is negative, so the fixed-area saddle, unlike the fixed-$\Lambda$ saddle of section~\ref{sec:saddlesol}, is the antipodal quotient of a real spindle of positive curvature $K=-4\Lambda_{\rm eff}=2\pi\gc/A$. This spindle has area $2A$ and two antipodal conical singularities of cone angle $2\pi\gc$. At the cone tip the finite part of the field is
\begin{equation}\label{eq:fa_conetip}
\varphi_*(0)\big|_A=-2\log c=\log\frac{2\gc A}{\pi}.
\end{equation}

We evaluate the classical action of this saddle in appendix~\ref{app:fa} in the same way as before. The only difference is the absence of the cosmological-constant term, which contributed $\gc/2$ to \eqref{eq:Scl-assemble}, so that
\begin{equation}\label{eq:fa_assemble}
-S_{\mathrm{cl}}(A)=\eta\,\varphi_*(0)-\mathcal{S}^{(0)}[\varphi_*]=\gc\log c+\frac{\gc}{2},
\end{equation}
and, with $\log c$ given by \eqref{eq:fa_conetip},
\begin{equation}\label{eq:fa_result}
S_{\mathrm{cl}}(A)=\frac{\gc}{2}\log A+\frac{\gc}{2}\Big(\log\frac{2\gc}{\pi}-1\Big)
\qquad(\gc>0,\ A>0),
\end{equation}
which is real. Thus the classical weight in the area integral is $\ee^{-S_{\mathrm{cl}}(A)/b^2}=A^{-\nu}$ up to an $A$-independent factor. Since $\nu=\gc/2b^2$ is large and positive, this weight is not integrable at $A=0$. The area integral, which also contains the one-loop factors of section~\ref{sec:oneloop}, therefore diverges at small $A$. We define it by analytic continuation in section~\ref{sec:zeromode}.

\section{One-loop determinant}
\label{sec:oneloop}

As is standard in Liouville theory, we separate the zero mode of $\varphi$ from the other fluctuations by working at fixed area. Inserting $1=\int_0^\infty\dd A\,\delta(\int_{r\le1}\ee^{\varphi}-A)$ into the path integral and extracting the cosmological-constant term, which equals $\mu A$ on the constraint surface, we obtain
\begin{align}
U(\alpha) &= \int_0^\infty \dd A\,\ee^{-\mu A}\,Z(\eta;A),\label{Ltranform}\\
Z(\eta;A) &= \int\mathcal{D}\varphi\;\ee^{\eta\varphi(0)/b^2}\,\ee^{-\mathcal{S}^{(0)}[\varphi]/b^2}\,\ee^{-S_{\rm bg}[\varphi]}\,
\delta\!\Big(\!\int_{r\le1}\!\dd^2z\,\ee^{\varphi}-A\Big)\notag\\
&\simeq \ee^{-S_{\mathrm{cl}}(A)/b^2}\,\ee^{-S_{\rm bg}}\,(\Det D)^{-1/2}\,\big(2\pi\,\mathcal{J}(A)\big)^{-1/2}.
\end{align}
Here $\mathcal{S}^{(0)}$ is the action \eqref{eq:reduced} without the cosmological-constant term, $S_{\mathrm{cl}}(A)$ is the fixed-area action \eqref{eq:fa_result}, $S_{\rm bg}$ is the $O(b^0)$ part of the background-charge coupling, $D$ is the fluctuation operator, and $\mathcal{J}(A)=\langle\ee^{\varphi_*},D^{-1}\ee^{\varphi_*}\rangle$ is the Jacobian produced by the delta function, which imposes a single linear condition on the fluctuation. In this section $\langle f,g\rangle=\int_{r\le1}\dd^2z\,fg$ denotes the flat inner product on the fundamental domain and $\Delta_\delta$ denotes the flat Laplacian.

The advantage of working at fixed area is that $\Det D$ does not depend on $A$. With $\varphi=\varphi_*+\xi$, the fluctuation operator is $D=-\Delta_\delta+8\Lambda\,\ee^{\varphi_*}$. The factors of $\Lambda$ cancel in the potential,
\begin{equation}\label{eq:potential}
8\Lambda\,\ee^{\varphi_*}=-\frac{2\gc^2}{r^2\cosh^2(\gc\log r)},
\end{equation}
which depends only on the cone angle. At fixed area the same potential arises from the area constraint, with $\Lambda_{\rm eff}$ in place of $\Lambda$.\footnote{\label{fn:fapotential}Since $\mathcal{S}^{(0)}$ has no cosmological-constant term, its second variation is just $-\Delta_\delta$, so one might expect $D$ to have no potential term. The potential arises instead from the area constraint. The fixed-area saddle extremizes $\eta\varphi(0)-\mathcal{S}^{(0)}$ only subject to the constraint, so it is not an unconstrained critical point. By the on-shell relation of appendix~\ref{app:fa} the gradient at the saddle is $\tfrac{\Lambda_{\rm eff}}{\pi}\ee^{\varphi_*}$, so a linear term $\tfrac{\Lambda_{\rm eff}}{\pi b^{2}}\langle\ee^{\varphi_*},\xi\rangle$ survives in the exponent. Once the constraint is expanded to second order, $\int\ee^{\varphi}-A=\langle\ee^{\varphi_*},\xi\rangle+\tfrac12\int\ee^{\varphi_*}\xi^{2}+O(\xi^{3})$, the delta function sets $\langle\ee^{\varphi_*},\xi\rangle=-\tfrac12\int\ee^{\varphi_*}\xi^{2}$ on its support. The surviving linear term then becomes the quadratic term $-\tfrac{\Lambda_{\rm eff}}{2\pi b^{2}}\int\ee^{\varphi_*}\xi^{2}$, which gives the potential $8\Lambda_{\rm eff}\ee^{\varphi_*}$ in $D$. The linear part $\delta(\langle\ee^{\varphi_*},\xi\rangle)$ of the constraint is treated in section~\ref{sec:jacobian}. The cubic and higher terms in $\xi$ contribute only beyond one loop.} Since $\Lambda_{\rm eff}=-\pi\gc/2A$ and $\ee^{\varphi_*}\propto A$, the operator $D$ is independent of $A$. Because $\Det D$ is defined with respect to the flat fiducial metric, the conformal anomaly does not introduce any $A$-dependence. Thus the integrand depends on $A$ only through $S_{\mathrm{cl}}(A)$, $\mathcal{J}(A)$, and $\ee^{-S_{\rm bg}}$.

\subsection{Evaluation of \texorpdfstring{$\Det D$}{Det D}}
\label{sec:detD}

The determinant factorizes over angular momenta. In the mode expansion $\xi=\sum_{m\in\mathbb Z}\psi_m(r)\,\ee^{im\theta}$ the operator acts on each mode as
\begin{equation}\label{eq:Dm}
D_m=-\partial_r^2-\frac1r\,\partial_r+\frac{m^2}{r^2}
-\frac{2\gc^2}{r^2\cosh^2(\gc\log r)} ,
\end{equation}
which depends on $m$ only through $m^2$, so the modes $\pm m$ are degenerate. At the singular endpoint $r=0$ regularity requires $\psi_m\sim r^{|m|}$. At $r=1$ the identification $\xi(\sigma z)=\xi(z)$ requires $\psi_m(1/r)=(-1)^m\psi_m(r)$, and evaluating this relation and its derivative at $r=1$ gives a Neumann condition for even $m$ and a Dirichlet condition for odd $m$,
\begin{equation}\label{eq:bc}
\psi_m'(1)=0\ \ (m\ \text{even}),\qquad \psi_m(1)=0\ \ (m\ \text{odd}).
\end{equation}

We evaluate each $\det D_m$ with the Gelfand--Yaglom theorem~\cite{GelfandYaglom,Dunne} in the form given in appendix~\ref{app:gy}. The theorem expresses the determinant in terms of the boundary value at $r=1$ of the zero-energy solution regular at $r=0$, without reference to the spectrum. In the variable $x=\gc\log r$ the equation $D_m\psi=0$ is the reflectionless P\"oschl--Teller equation $\psi_{xx}-(m^2/\gc^2)\psi+2\,\sech^2x\,\psi=0$, whose regular solution is
\begin{equation}\label{eq:reg-sol}
f_m(r)=r^{|m|}\Big(\tanh(\gc\log r)-\frac{|m|}{\gc}\Big),
\end{equation}
with $f_m(1)=-|m|/\gc$ and $f_m'(1)=(\gc^2-m^2)/\gc$. For $m\neq0$ we normalize the determinant by that of the free operator $D_m^{0}$, which is $D_m$ without the potential and has the regular solution $r^{|m|}$. Once $f_m$ is normalized to the same leading behavior $r^{|m|}$ as $r\to0$, which amounts to dividing it by $-(\gc+|m|)/\gc$, the boundary conditions \eqref{eq:bc} give
\begin{equation}\label{eq:per-mode}
\frac{\det D_m}{\det D_m^{0}}=
\begin{cases}
\dfrac{|m|-\gc}{|m|}, & m\ \text{even},\\[1.2ex]
\dfrac{|m|}{|m|+\gc}, & m\ \text{odd}.
\end{cases}
\end{equation}
Using the degeneracy of $\pm m$ and writing $m=2k$ for even and $m=2k-1$ for odd modes, we obtain
\begin{equation}\label{eq:sums}
\log\frac{\Det D}{\Det D^{0}}\bigg|_{m\neq0}
=2\sum_{k\ge1}\log\frac{2k-\gc}{2k}
+2\sum_{k\ge1}\log\frac{2k-1}{2k-1+\gc}.
\end{equation}

Each series diverges logarithmically. With an angular-momentum cutoff $M$, the two partial sums together grow as $-2\gc\log M$. This divergence is linear in $\eta$, so we can absorb it into a multiplicative renormalization of the cosmological constant. We compute the finite parts by zeta regularization, using the identity $\sum_{k\ge0}[\log(k+a_1)-\log(k+a_2)]_{\rm reg}=\log[\Gamma(a_2)/\Gamma(a_1)]$, a consequence of Lerch's formula for the Hurwitz zeta function. The sum over even $m$ gives $-2\log\Gamma(1-\tfrac\gc2)$, and the sum over odd $m$ gives $2\log\!\big[\Gamma(\tfrac{1+\gc}2)/\Gamma(\tfrac12)\big]$. The duplication and reflection formulas for the $\Gamma$-function combine them into
\begin{equation}\label{eq:mneq0}
-\tfrac12\log\frac{\Det D}{\Det D^{0}}\bigg|_{m\neq0}
=\log\Gamma(2\eta)+\log\sin\pi\eta+\gc\log2 .
\end{equation}

Thus the transcendental factors $\Gamma(2\eta)$ and $\sin\pi\eta$ of \eqref{eq:nuneg_W1} are produced by the modes with $m\neq0$, and no $\log\gc$ arises in this sector. The remaining term $\gc\log2$ is linear in $\eta$ and depends on the choice of the reference operator $D_m^{0}$ and on how the divergent sums \eqref{eq:sums} are regularized. Such an $\eta$-linear term can be absorbed into a finite renormalization of the cosmological constant. We fix its value in appendix~\ref{app:renorm}, where we evaluate $\Det D$ in the renormalization scheme that matches the normalization of the cosmological constant in the exact formula. The same evaluation confirms that the factors $\Gamma(2\eta)$ and $\sin\pi\eta$ do not depend on the renormalization scheme.

The $m=0$ sector is even, so its determinant is given by the Neumann boundary value. This determinant cannot be normalized by that of the free operator, whose regular solution is the constant function and has vanishing Neumann boundary value. The operator $D_0$ itself, however, has no zero mode because its regular solution $f_0(r)=\tanh(\gc\log r)$ has $f_0'(1)=\gc\neq0$ and does not satisfy the Neumann condition. The Gelfand--Yaglom theorem then gives $\det D_0\propto\gc$ directly from this boundary value,
\begin{equation}\label{eq:m0}
-\tfrac12\log\det D_0=-\tfrac12\log\gc+\text{const}.
\end{equation}
The undetermined constant comes from the regularization at the singular endpoint $r\to0$. It is independent of $\eta$, since $f_0$ approaches the $\eta$-independent value $f_0(0^+)=-1$ there. It therefore affects only the overall normalization. Altogether,
\begin{equation}\label{eq:detD}
-\tfrac12\log\Det D=\log\Gamma(2\eta)+\log\sin\pi\eta+\gc\log2-\tfrac12\log\gc+\text{const},
\end{equation}
with an $\eta$-independent constant that also absorbs the determinant $\Det D^0|_{m\neq0}$ of the free reference operator.

\subsection{Evaluation of \texorpdfstring{$\mathcal{J}(A)$}{J(A)}}
\label{sec:jacobian}

With $\varphi=\varphi_*+\xi$ and $\int\ee^{\varphi_*}=A$, the area constraint reduces to $\delta(\langle\ee^{\varphi_*},\xi\rangle)$, since its quadratic part has been absorbed into the potential of $D$ (see footnote~\ref{fn:fapotential}). The constrained Gaussian integral is
\begin{equation}\label{eq:constrained-gaussian}
\int\mathcal{D}\xi\;\delta\!\big(\langle\ee^{\varphi_*},\xi\rangle\big)\,
\ee^{-\frac12\langle\xi,D\xi\rangle}
=(\Det D)^{-1/2}\,(2\pi\,\mathcal{J})^{-1/2},
\qquad \mathcal{J}=\langle\ee^{\varphi_*},D^{-1}\ee^{\varphi_*}\rangle,
\end{equation}
where $\Det D$ is the full determinant of section~\ref{sec:detD} and the Jacobian $\mathcal{J}$ is finite. Both $\ee^{\varphi_*}$ and $D^{-1}\ee^{\varphi_*}$ lie in the $m=0$ sector, where $D_0\,u=\ee^{\varphi_*}$ is solved by the constant function $u=1/8\Lambda_{\rm eff}$, so that
\begin{equation}\label{eq:J}
\mathcal{J}(A)=\frac1{8\Lambda_{\rm eff}}\int\ee^{\varphi_*}=-\frac{A^2}{4\pi\gc},
\qquad
\big(2\pi|\mathcal{J}|\big)^{-1/2}=\frac{\sqrt{2\gc}}{A}.
\end{equation}
The negative sign of $\mathcal{J}$ comes from the constant mode, along which $\langle 1,D\,1\rangle=8\Lambda_{\rm eff}A<0$ and an unconstrained Gaussian integral would diverge. In fact $D$ has exactly one negative mode. For $m\neq0$, in the variable $x=\gc\log r$, the operator $r^2D_m=\gc^2\big(-\partial_x^2-2\,\sech^2x+m^2/\gc^2\big)$ is positive, since the lowest eigenvalue of $-\partial_x^2-2\,\sech^2x$ is $-1$ and $m^2/\gc^2>1$. For $m=0$, the regular solution $f_0=\tanh(\gc\log r)$ of $D_0f_0=0$ has no node in $0<r<1$ and vanishes at $r=1$, so zero is the lowest Dirichlet eigenvalue of $D_0$, and exactly one Neumann eigenvalue lies below it. Hence $\Det D<0$, and since $\mathcal{J}<0$ as well, $D$ is positive on the constraint surface $\langle\ee^{\varphi_*},\xi\rangle=0$. Fixing the area therefore removes the instability, and the phases of $(\Det D)^{-1/2}$ and $\mathcal{J}^{-1/2}$ cancel, so the constrained Gaussian integral is real and positive. The absolute value $(2\pi|\mathcal{J}|)^{-1/2}$ contributes $+\tfrac12\log\gc$ to the one-loop term and cancels the $-\tfrac12\log\gc$ of the $m=0$ determinant \eqref{eq:m0}. Unlike $\Det D$, the Jacobian factor $(2\pi|\mathcal{J}|)^{-1/2}\propto A^{-1}$ depends on the area, so it belongs to the integrand of the area integral in section~\ref{sec:zeromode}.

\section{The zero-mode integral and the sum over saddles}
\label{sec:zeromode}

It remains to integrate over the area $A$. Apart from $\ee^{-\mu A}$, the integrand is a product of powers of $A$ coming from the classical action, the Jacobian, and the background-charge coupling. For $0<\eta<\tfrac12$ the integral diverges at small $A$ and has to be defined by analytic continuation. This integral and the renormalization scheme fixed in appendix~\ref{app:renorm} give the remaining $\eta$-dependence of \eqref{eq:nuneg_W1}. When the integral is written in terms of the zero mode, its analytic continuation becomes the sum over the complex saddles of section~\ref{sec:saddle}.

\subsection{The area integral and its semiclassical expansion}
\label{sec:assemble}

We now evaluate the $O(b^0)$ part $S_{\rm bg}=\tfrac1{8\pi}\int\sqrt{\hat g}\,\hat R\,\varphi_*$ of the background-charge coupling. The curvature is supported on the circle $r=1$, on which $\varphi_*$ is constant. Since $\tfrac1{4\pi}\int\sqrt{\hat g}\,\hat R=\chi(\mathbb{RP}^2)=1$, we have $S_{\rm bg}=\tfrac12\varphi_*(1)$. On the fixed-area saddle, \eqref{eq:fa_conetip} and $g(1)=2$ give $\varphi_*(1)=\log(\gc A/2\pi)$. Hence
\begin{equation}\label{eq:bgcharge}
\ee^{-S_{\rm bg}}=A^{-1/2}\,\gc^{-1/2}\,(2\pi)^{1/2}. \end{equation}

We can now collect the area dependence of the fixed-area integrand. The classical action \eqref{eq:fa_result} gives $A^{-\nu}$, the Jacobian \eqref{eq:J} gives $A^{-1}$, the background-charge coupling \eqref{eq:bgcharge} gives $A^{-1/2}$, and $\Det D$ is independent of $A$. Their product is $A^{-\nu-3/2}=A^{-s-1}$ with $s=\nu+\tfrac12$, and the area integral becomes
\begin{equation}\label{eq:Aint}
\int_0^\infty \dd A\,A^{-s-1}\ee^{-\mu A}=\Gamma(-s)\,\mu^{s} .
\end{equation}
The power of $\mu$ is $\nu+\tfrac12$, in agreement with the exponent $(Q-2\alpha)/2b$ in the exact formula \eqref{eq:U}. The half-integer shift away from the classical value $\nu$ comes from the background-charge coupling, that is, from the Euler characteristic $\chi(\mathbb{RP}^2)=1$. As a function of $s$, the integral \eqref{eq:Aint} converges only for $\operatorname{Re}s<0$, and $\Gamma(-s)\mu^{s}$ is its analytic continuation to the physical region $s=\nu+\tfrac12>0$, where the integral diverges as $A\to0$. We define the integral by this analytic continuation, known as the Goulian--Li prescription~\cite{GoulianLi}. In section~\ref{sec:contour} we realize it as a contour deformation.

Collecting all factors, we find
\begin{equation}\label{eq:Ufull}
U=\ee^{-C/b^2}\,\Gamma(2\eta)\,\sin\pi\eta\;2^{\gc}\,\gc^{-1/2}\,\Gamma(-s)\,\mu^{s}\,\ee^{W_0},
\qquad
C=\frac{\gc}{2}\Big(\log\frac{2\gc}{\pi}-1\Big),\quad \mu=\frac{\Lambda}{\pi b^2},
\end{equation}
where, as in \eqref{eq:nuneg_exp}, $W_0$ stands for $\eta$-independent terms that we do not keep track of. In \eqref{eq:Ufull} it contains the overall normalization, which is real and $b$-dependent.\footnote{In the exact formula~\eqref{eq:U} the overall normalization is fixed by the modular bootstrap applied to the M\"obius-strip amplitude between the ZZ brane and the crosscap state. We do not determine $W_0$ from the path integral in this paper.} By the reflection formula $\Gamma(-s)=-\pi\big[\Gamma(\nu+\tfrac32)\cos\pi\nu\big]^{-1}$, the factor $\Gamma(-s)$ splits into an oscillating factor $1/\cos\pi\nu$ and a smooth factor $\Gamma(\nu+\tfrac32)^{-1}$. With $\nu=\gc/2b^2$, $\mu=\Lambda/\pi b^2$, and the Stirling expansion $\log\Gamma(\nu+\tfrac32)=(\nu+1)\log\nu-\nu+\tfrac12\log2\pi+O(\nu^{-1})$, the smooth part is
\begin{align}\label{eq:logU-two}
\log\big|U\cos\pi\nu\big|
&=\frac1{b^2}\Big(\frac{\gc}{2}\log\Lambda-\gc\log\gc+\gc\Big)\notag\\
&\quad+\Big(\log\Gamma(2\eta)+\log\sin\pi\eta-\tfrac32\log\gc+\gc\log2\Big)+W_0+O(b^2).
\end{align}

We now compare \eqref{eq:logU-two} with the semiclassical limit of section~\ref{sec:setup}. The leading term is the classical action \eqref{eq:nuneg_action}, and three of the four one-loop terms, $\log\Gamma(2\eta)$, $\log\sin\pi\eta$, and $-\tfrac32\log\gc$, coincide with the corresponding terms in \eqref{eq:nuneg_W1}. The fourth term, $\gc\log2$, depends on the renormalization scheme. With the renormalization scheme fixed in appendix~\ref{app:renorm} by matching the normalization of the cosmological constant in the exact formula, it becomes $-\gc\gamma_E$, and \eqref{eq:logU-two} reproduces the full $\eta$-dependence of \eqref{eq:nuneg_W1}. We discuss the factor $1/\cos\pi\nu$ in the next two subsections.

\subsection{Deforming the \texorpdfstring{$A$}{A}-integration contour}
\label{sec:contour}

\begin{figure}[tb]
\centering
\resizebox{\textwidth}{!}{%
\begin{tikzpicture}[>=Stealth]

\begin{scope}
  \node[anchor=west,font=\bfseries] at (-3.2,3.1) {(a) $A$-plane};
  \draw[axisline] (-3.2,0) -- (4.0,0) node[below right,black]{$\operatorname{Re}A$};
  \draw[axisline] (0,-3.0) -- (0,3.0) node[right,black]{$\operatorname{Im}A$};
  \draw[cutline] (0.05,0) -- (3.7,0);
  \node[brown!70!black] at (3.0,-0.8) {\small cut};
  \draw[contour]
     (3.7,0.42) -- (0.7,0.42)
     .. controls (0.0,0.42) and (-0.55,0.22) .. (-0.55,0)
     .. controls (-0.55,-0.22) and (0.0,-0.42) .. (0.7,-0.42)
     -- (3.7,-0.42);
  \node[blue!70!black] at (2.2,0.85) {$\mathcal{H}$};
  \node[saddle] at (-2.0,0) {};
  \node[anchor=west,black,fill=white,inner sep=1.2pt] at (-1.82,-0.42) {$A_*=-\dfrac{s}{\mu}$};
  \node[black,align=center,fill=white,inner sep=1.2pt] at (-1.10,1.35) {\small single\\[-1pt]\small saddle};
  \draw[sdpath] (-2.0,-2.5) -- (-2.0,2.5);
  \node[red!75!black,rotate=90,anchor=south,font=\small,fill=white,inner sep=1.2pt] at (-2.28,-1.75) {steepest descent};
  \draw[->,gray,thick,dashed] (0.45,1.85) .. controls (-0.9,2.25) .. (-1.78,1.95);
  \node[gray,font=\small] at (-0.5,2.4) {deform};
\end{scope}

\begin{scope}[shift={(6.25,0)}]
  \draw[-{Stealth[length=3mm]},line width=1.1pt] (-0.85,0.15) -- (0.85,0.15);
  \node[above] at (0,0.28) {$A=\ee^{\varphi_0}$};
  \node[below,gray,font=\small] at (0,0.02) {$\varphi_0=\log A$};
\end{scope}

\begin{scope}[shift={(11.0,0)}]
  \node[anchor=west,font=\bfseries] at (-3.0,3.1) {(b) $\varphi_0$-plane};
  \draw[axisline] (-3.0,0) -- (3.6,0) node[below right,black]{$\operatorname{Re}\varphi_0$};
  \draw[axisline] (0,-4.9) -- (0,3.0) node[right,black]{$\operatorname{Im}\varphi_0$};
  \foreach \y in {2,0,-2,-4}{
     \draw[gray!55,dashed,thin] (-3.0,\y) -- (3.4,\y);
  }
  \node[gray,font=\small,fill=white,inner sep=1.2pt] at (-2.35,-0.4) {$A\to0$};
  \node[gray,font=\small,fill=white,inner sep=1.2pt] at (2.7,0.42) {$A\to\infty$};
  \node[ghost] at (1.5,1) {};
  \node[right,font=\small,gray!60!black,fill=white,inner sep=1.2pt] at (1.7,1) {$\varphi_0^{*(0)}$};
  \node[left,gray!60!black,font=\scriptsize,fill=white,inner sep=1.2pt] at (1.30,1) {not crossed};
  \node[gray!55,font=\small] at (1.5,2.35) {$\vdots$};
  \node[saddle] at (1.5,-1) {};
  \node[saddle] at (1.5,-3) {};
  \node[right,font=\small,fill=white,inner sep=1.2pt] at (1.7,-1) {$\varphi_0^{*(1)}$};
  \node[right,font=\small,fill=white,inner sep=1.2pt] at (1.7,-3) {$\varphi_0^{*(2)}$};
  \node at (1.5,-4.75) {$\vdots$};
  \draw[decorate,decoration={brace,amplitude=4pt},gray!85!black]
       (1.15,-1.0) -- (1.15,-3.0);
  \node[gray!85!black,left,align=center,font=\small,fill=white,inner sep=1.2pt] at (1.0,-2.0) {shift\\ $-2\pi i$};
  \draw[contour]
     (3.4,0.0) .. controls (2.3,0.0) and (1.5,-0.35) .. (1.5,-1.0)
     -- (1.5,-4.35);
\end{scope}

\node[align=center,black] at (0.4,-3.7)
   {$\Rightarrow\ \Gamma\!\big(\nu+\tfrac32\big)^{-1}$ \ (smooth)};
\node[align=center,black] at (11.0,-5.35)
   {$\Rightarrow\ 1/\cos\pi\nu$ \ (oscillation)};

\end{tikzpicture}%
}
\caption{Contour structure of the zero-mode integral, discussed in sections~\ref{sec:contour}--\ref{sec:phicontour}. (a)~The $A$-plane. The integrand $A^{-s-1}\ee^{-\mu A}$ has a branch cut along the positive real axis, and the Hankel contour $\mathcal{H}$ of \eqref{eq:hankel} is deformed onto the steepest-descent path through the stationary point $A_*=-s/\mu$. (b)~The zero-mode integral \eqref{eq:zeromode-int}, with $A=\ee^{\varphi_0}$, in which $A_*$ corresponds to the family $\varphi_0^{*(n)}=\varphi_0^{*(0)}-2\pi i\,n$. The dashed horizontal lines are the images of $A\in(0,\infty)$ on the successive sheets. The steepest-descent contour passes through the saddles with $n\ge1$ but not through those with $n\le0$, such as $\varphi_0^{*(0)}$ (open circle).}
\label{fig:contour}
\end{figure}
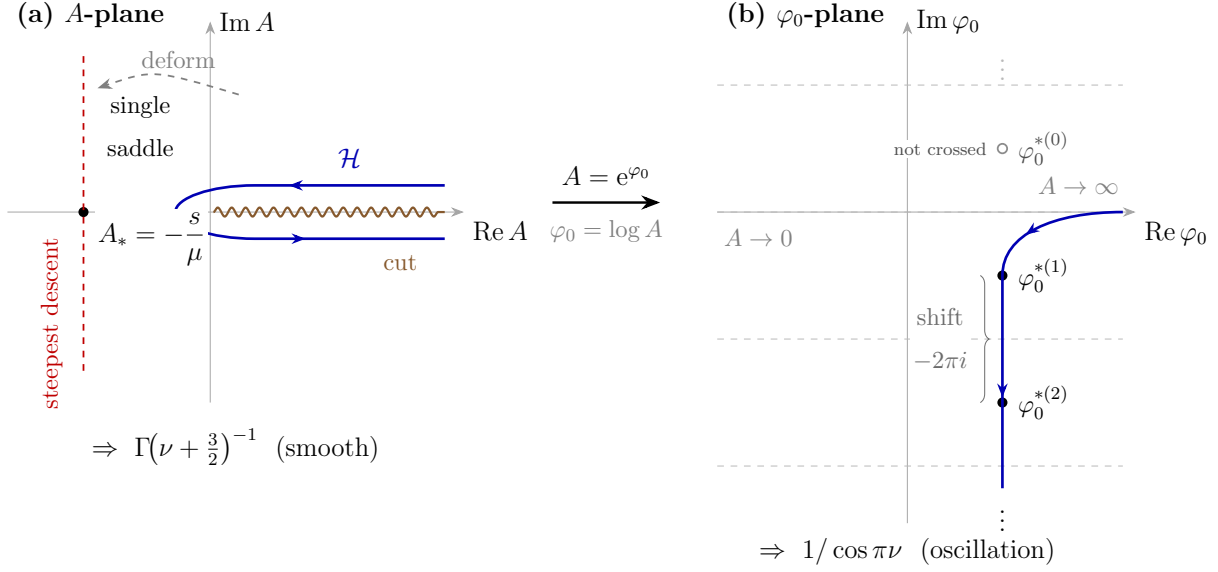

We now implement the analytic continuation as a contour deformation (figure~\ref{fig:contour}(a)). The integrand $A^{-s-1}\ee^{-\mu A}$ has a branch point at $A=0$. We take the Hankel contour $\mathcal{H}$ that comes in from $+\infty$ above the positive real axis, goes around the origin, and returns to $+\infty$ below it. Then
\begin{equation}\label{eq:hankel}
\Gamma(-s)\,\mu^{s}=\frac{1}{\ee^{-2\pi i s}-1}\oint_{\mathcal{H}}\dd A\;A^{-s-1}\,\ee^{-\mu A}.
\end{equation}
For $\operatorname{Re}s<0$ the two edges of the branch cut give $(\ee^{-2\pi i s}-1)$ times the convergent integral \eqref{eq:Aint}. For $s=\nu+\tfrac12>0$ the integral \eqref{eq:Aint} diverges as $A\to0$, but the contour integral on the right-hand side of \eqref{eq:hankel} is an entire function of $s$ because $\mathcal{H}$ avoids the origin. Therefore the right-hand side of \eqref{eq:hankel} gives the analytic continuation. As the constant mode of $\varphi$ runs over the real line, $A=\int_{r\le1}\ee^{\varphi}\,\dd^2z$ ranges over $(0,\infty)$, so $\mathcal{H}$ can also be regarded as a contour for the zero mode.

In the semiclassical limit, where $s$ and $\mu$ are both $O(b^{-2})$, the integrand behaves as $\ee^{-s\log A-\mu A}$ and has a single stationary point
\begin{equation}\label{eq:Asaddle}
A_*=-\frac{s}{\mu}
\end{equation}
on the negative real axis. We can deform $\mathcal{H}$ to the steepest-descent contour through it. Thus the dominant area is negative. The stationary point $A_*$ corresponds to the saddle of section~\ref{sec:saddle}. As $b\to0$ it approaches $-\pi\gc/2\Lambda$, the area $\int_{r\le1}\ee^{\varphi_*}\dd^2z$ of the fixed-$\Lambda$ saddle, which is negative because the Weyl factor \eqref{eq:metric} is negative. Evaluating \eqref{eq:fa_Leff} at \eqref{eq:Asaddle} likewise gives
\begin{equation}\label{eq:csq-recovered}
c^2=\frac{\pi}{2A_*\gc}\;\xrightarrow{\ b\to0\ }\;-\frac{\Lambda}{\gc^2},
\end{equation}
which is the condition \eqref{eq:csq} for the fixed-$\Lambda$ saddle. The steepest-descent evaluation at this saddle gives the contour integral in \eqref{eq:hankel}. This integral has no poles and reproduces, up to a phase, the smooth factor $\Gamma(\nu+\tfrac32)^{-1}$ of \eqref{eq:logU-two}.\footnote{Explicitly, since $\ee^{-2\pi i s}-1=-2i\,\ee^{-i\pi s}\sin\pi s$, multiplying $\Gamma(-s)\mu^{s}$ by this factor removes the $1/\sin\pi s=1/\cos\pi\nu$, so that $\oint_{\mathcal{H}}\dd A\,A^{-s-1}\ee^{-\mu A}=2\pi i\,\ee^{-i\pi s}\mu^{s}/\Gamma(s+1)$ has no poles. Its phase $\ee^{-i\pi s}$ cancels against that of the prefactor.} Where do the oscillations come from? The area does not distinguish the members of the family of section~\ref{sec:saddle}, since they all have the same Weyl factor and hence the same area $A_*$. The oscillations come instead from the prefactor $1/(\ee^{-2\pi i s}-1)$ in \eqref{eq:hankel}, which is due to the monodromy of $A^{-s-1}$ around the branch point. In the $A$-plane this prefactor is just a number multiplying a contour integral without poles. It becomes a sum over saddles only when the integral is written in terms of the zero mode.

\subsection{The family of complex saddles from the zero mode}
\label{sec:phicontour}

We set $A=\ee^{\varphi_0}$, where $\varphi_0$ is the constant mode of $\varphi$ and the shape-dependent part of the area has been absorbed into an additive shift of $\varphi_0$. In this variable \eqref{eq:Aint} reads
\begin{equation}\label{eq:zeromode-int}
\int_0^\infty\dd A\,A^{-s-1}\ee^{-\mu A}=\int_{-\infty}^{\infty}\dd\varphi_0\;\ee^{-\mathcal{W}(\varphi_0)},
\qquad
\mathcal{W}(\varphi_0)=s\,\varphi_0+\mu\,\ee^{\varphi_0},
\end{equation}
where $\mathcal{W}$ is the effective action of the zero mode. Its stationary point $\ee^{\varphi_0^*}=-s/\mu$ is the stationary point \eqref{eq:Asaddle} of the area integral. However, $\varphi_0=\log A$ is multivalued, so this stationary point corresponds to the infinite family
\begin{equation}\label{eq:zeromode-tower}
\varphi_0^{*(n)}=\varphi_0^{*(0)}-2\pi i\,n,\qquad n\in\mathbb{Z}.
\end{equation}
Here $\varphi_0^{*(0)}=\log(s/\mu)+i\pi$. The points $\varphi_0^{*(n)}$ are the constant modes of the complex saddles \eqref{eq:phi-tower} of section~\ref{sec:saddle}, with the label shifted by one, so that $\varphi_0^{*(n)}$ is the constant mode of $\varphi_*^{(n-1)}$. They form the orbit of the stationary point $\varphi_0^{*(0)}$ under the deck transformation $\varphi_0\to\varphi_0-2\pi i$ of the logarithmic covering $A=\ee^{\varphi_0}$. This shift leaves $A$ fixed, so all of them lie over the point $A_*$ of the $A$-plane, one on each sheet of the covering (figure~\ref{fig:contour}(b)).

On the real $\varphi_0$-axis, the image of $A\in(0,\infty)$, the integral still diverges, now as $\varphi_0\to-\infty$. For $\operatorname{Im}\nu>0$, as in \eqref{eq:cosexp}, it converges on a contour that descends through the sheets from large real $\varphi_0$. This contour decomposes into steepest-descent paths, or Lefschetz thimbles, each passing once through one of the saddles with $n\ge1$.\footnote{The actions of the saddles are $\mathcal{W}(\varphi_0^{*(n)})=s[\log(s/\mu)-1]+i(1-2n)\pi s$. For $\operatorname{Im}\nu>0$ their real parts grow linearly with $n$, so the sum over the saddles with $n\ge1$ converges, while the sum over those with $n\le0$ diverges. For $\operatorname{Re}s\neq0$ their imaginary parts are all distinct, so no two saddles are connected by a steepest-descent path, and the thimbles are well defined. The physical case $\operatorname{Im}\nu=0$ is the anti-Stokes line, on which all saddles have the same real part of the action. In~\cite{HMW} the corresponding line for the two-point function on the sphere is called a Stokes line, since the set of contributing saddles changes across it.} The integral over the $n$-th thimble is $\ee^{-\mathcal{W}(\varphi_0^{*(n)})}$ times a Gaussian factor common to all $n$, since the Hessian $\mathcal{W}''=-s$ does not depend on $n$. The deck transformation shifts $\mathcal{W}$ by $-2\pi i s$, so consecutive contributions differ by the constant factor $\ee^{2\pi i s}$. With $s=\nu+\tfrac12$ the $n$-th contribution is, up to an $n$-independent factor, $\ee^{2\pi i n s}=(-1)^n\,\ee^{2\pi i n\nu}$. Summing over $n\ge1$ gives
\begin{equation}\label{eq:deck-sum}
\sum_{n\ge1}(-1)^n\,\ee^{2\pi i n\nu}=\frac{1}{\ee^{-2\pi i s}-1},
\end{equation}
which is the prefactor of \eqref{eq:hankel}. Term by term, the sum matches the expansion \eqref{eq:cosexp} with $n$ shifted by one, up to an $n$-independent factor. This factor combines with the contour integral of section~\ref{sec:contour} into the smooth part of the one-point function. The prefactor $1/(\ee^{-2\pi i s}-1)$, which had no saddle-point interpretation in the $A$-plane, is therefore the sum over the complex saddles of section~\ref{sec:saddle} with their relative phases.

To summarize, the stationary point of the area integral gives the smooth factor $\Gamma(\nu+\tfrac32)^{-1}$, and the family of complex saddles obtained from the multivalued zero mode gives the oscillating factor $1/\cos\pi\nu$. Together with the one-loop factors of section~\ref{sec:oneloop} and the renormalization scheme fixed in appendix~\ref{app:renorm}, these factors reproduce the semiclassical limit \eqref{eq:nuneg_exp} of the exact one-point function.

\section{Discussion}
\label{sec:summary}

We have shown that, through one-loop order, the crosscap one-point function of Liouville theory obtained from the bootstrap~\cite{Hikida,BergmanHirano} is reproduced by a semiclassical path integral around complex saddles whose metric is negative definite, provided the divergent integral over the zero mode is defined by analytic continuation.

The oscillation of the one-point function comes from this analytic continuation. For the integral over the area, the Goulian--Li prescription~\cite{GoulianLi} leads to the monodromy factor $1/(\ee^{-2\pi i s}-1)$, which has no saddle-point interpretation. For the integral over the zero mode, the same factor is a geometric series over saddles that differ by the shift $\varphi\to\varphi-2\pi i$. This mechanism was found in~\cite{HMW} for the two- and three-point functions on the sphere. There, too, the complex saddles differ only by shifts of the zero mode, and the integral over the zero mode is the integral representation of the $\Gamma$-function. For the sphere partition function of timelike Liouville theory, a Picard--Lefschetz analysis of a Hankel-type contour for the zero mode was recently shown to reproduce an oscillatory factor of the analytically continued DOZZ formula~\cite{Despontin}. Saddles shifted by $2\pi i n$ were also noted in the disk path integral of timelike Liouville theory~\cite{LiouvilleCosmology}, where it was left open how they could produce the phases in the bootstrap result for the disk one-point function~\cite{BautistaBawane} (see also~\cite{GiribetSivilotti}). It would be interesting to see whether this mechanism also accounts for these phases.

Our saddles are not allowable in the sense of~\cite{KontsevichSegal,WittenComplex}, since their Weyl factor \eqref{eq:metric} is negative. The same is true of the complex saddles that describe the sphere correlators when no real solution exists~\cite{HMW}. Yet these saddles are needed to reproduce the exact answers. In~\cite{WittenComplex} it was suggested that the integration cycle of the gravitational path integral should lie in the space of allowable metrics. In our example this requirement is violated, but only along the conformal zero mode. At fixed area the saddle is the antipodal quotient of a real spindle, on which we evaluate the fluctuation determinant and the Jacobian. Only the total area is analytically continued to negative values. When the metric itself is integrated over, allowability conditions may therefore have to be relaxed along the conformal zero mode. This is reminiscent of the conformal factor problem of Euclidean gravity, where the contour of the conformal mode has to be rotated~\cite{GibbonsHawkingPerry,Polchinski}.

A related question is which complex saddles contribute. For a gravitational path integral this is in general a difficult question that requires a Picard--Lefschetz analysis~\cite{WittenCS}. For three-dimensional gravity with a cosmological constant, complex saddles were analyzed using a holographic Liouville theory, with the contour fixed in a mini-superspace reduction~\cite{ComplexSaddlesdS3,ComplexSaddlesCS3,Semiclassical3dSaddlesLetter,Semiclassical3dSaddles}. In our problem the question is simpler, since at fixed area the remaining fluctuations are Gaussian around a real saddle. The choice of contour then reduces to that for a single integral over the zero mode, and the contributing saddles are fixed by requiring this integral to converge.

Our analysis assumed a rotationally symmetric saddle. There are no non-symmetric saddles close to the spindle, since the fluctuation operator is positive for $m\neq0$ (section~\ref{sec:jacobian}). Nor are there non-symmetric saddles with a real Weyl factor. By the Gauss--Bonnet theorem, such a Weyl factor must be negative. Reversing its sign gives a metric of constant positive curvature on the double cover with two conical singularities of angle $2\pi\gc<2\pi$. Such a metric is necessarily rotationally symmetric~\cite{Troyanov1989} (see~\cite{Troyanov} for the general problem of prescribing curvature with conical singularities). However, these arguments do not exclude complex saddles far from the spindle. The exact answer nevertheless rules out contributions from additional saddles whose action has a larger real part than that of our saddles. Such contributions would produce singularities of the Borel transform in the right half-plane, whereas the only divergent series in the exact answer is the Stirling series of $\log\Gamma(\gc/b^2+2)$ in \eqref{eq:Gf}. Its Borel singularities lie only on the imaginary axis, at $\pm2\pi i k\gc$, and its Borel sum reproduces $\log\Gamma$ exactly.

It would also be interesting to extend our analysis to the Klein bottle, to amplitudes with both a crosscap and FZZT or ZZ boundaries, and to the bulk two-point function on $\mathbb{RP}^2$, whose crossing symmetry was studied in~\cite{FioravantiPradisiSagnotti,PradisiSagnottiStanevWZW,PradisiSagnottiStanev}. Since timelike Liouville theory has been used as a model of two-dimensional de~Sitter space~\cite{AnninosTwoSphere,AnninosBaraccoMuhlmann}, the timelike version of the crosscap one-point function may be relevant to two-dimensional elliptic de~Sitter space, whose analytic continuation to Euclidean signature is $\mathbb{RP}^2$~\cite{ParikhSavonijeVerlinde}. See also~\cite{WeiXCFT,LiXieHe} for holographic duals of crosscap conformal field theories involving de~Sitter end-of-the-world branes. Finally, the complex Liouville string~\cite{ComplexLiouvilleString} has recently been studied on crosscap geometries in connection with the double-scaled SYK model and three-dimensional de~Sitter space~\cite{BlommaertTiettoVerlinde2025,BlommaertTiettoVerlinde2026}. The unoriented complex Liouville string would presumably be dual to a matrix integral of orthogonal or symplectic type. Constructing this string theory would require the crosscap one-point function at complex central charge, together with contour prescriptions of the kind tested here.

\section*{Acknowledgments}

The author thanks Y. Wang for his continued encouragement to work on Liouville theory again. The author also thanks K. Shinmyo for discussions. This work is supported in part by JSPS KAKENHI Grant Numbers 21K03581 and 26K00699.

A large language model, Claude Fable 5.1 (Anthropic), was used for English language editing and for independent checks of intermediate formulas.

\appendix
\section{Evaluation of the classical action}\label{app:action}

In this appendix we evaluate the four terms in \eqref{eq:terms} and combine them into the classical actions \eqref{eq:Scl-assemble} and \eqref{eq:fa_result}, first at fixed cosmological constant and then at fixed area. In both cases we regulate the insertion at $r=\epsilon$ and integrate over the annulus $\epsilon\le r\le1$, with $\dd^2z=r\,\dd r\,\dd\theta$.

\subsection{Saddle at fixed cosmological constant}

We write the saddle as
\begin{equation}\label{eq:app-phi}
\varphi_*(r)=-2\log c-2\log g(r),\qquad
g(r)=r^{1+\gc}+r^{1-\gc},\qquad c^2=-\frac{\Lambda}{\gc^2},
\end{equation}
where $\log c=\tfrac12\log\Lambda-\log\gc+\tfrac{i\pi}2(2n+1)$ on the $n$-th branch. The branch label appears only in the additive constant $-2\log c$, since $\ee^{\varphi_*}=(c\,g)^{-2}=-\tfrac{\gc^2}{\Lambda}\,g^{-2}$ depends on $c$ only through $c^2$.

The source term is $\eta\,\varphi_*(\epsilon)$. Near the insertion $g(\epsilon)=\epsilon^{1-\gc}(1+\epsilon^{2\gc})$, so that $\varphi_*(\epsilon)=-2\log c-4\eta\log\epsilon+O(\epsilon^{2\gc})$ and
\begin{equation}
\eta\,\varphi_*(\epsilon)=-2\eta\log c-4\eta^2\log\epsilon+O(\epsilon^{2\gc}) .
\end{equation}

The kinetic term is
\begin{equation}
\frac1{4\pi}\int\dd^2z\,\partial\varphi_*\bar\partial\varphi_* =\frac12\int_\epsilon^1\frac{r\,g'^2}{g^2}\,\dd r ,
\end{equation}
with $\varphi_*'=-2g'/g$ and $g'=(1+\gc)r^\gc+(1-\gc)r^{-\gc}$. An integration by parts would leave a boundary term proportional to $\varphi_*(1)\varphi_*'(1)$, which does not vanish since $\varphi_*'(1)=-2$ by \eqref{eq:equiv}. We therefore evaluate the integral directly, using the variable $w=r^{2\gc}$, in which the integrand is rational,
\begin{equation}
\frac12\int_\epsilon^1\frac{r\,g'^2}{g^2}\,\dd r
=\frac1{4\gc}\int_{\epsilon^{2\gc}}^{1}
\frac{\big[(1+\gc)w+(1-\gc)\big]^2}{w\,(w+1)^2}\,\dd w
=-2\eta^2\log\epsilon+\log2-\frac\gc2+O(\epsilon^{2\gc}) .
\end{equation}
The divergence $-2\eta^2\log\epsilon$ comes from the $1/w$ part of the integrand.

The cosmological-constant term is finite as $\epsilon\to0$. In the variable $u=r^{\gc}$ we find
\begin{equation}
\frac{\Lambda}{\pi}\int\dd^2z\,\ee^{\varphi_*} =-2\gc\int_{0}^{1}\frac{u\,\dd u}{(u^2+1)^2}+O(\epsilon^{2\gc})=-\frac\gc2 .
\end{equation}

The curvature term is localized on the circle $r=1$, where the flat fiducial metric is not smooth. There the curvature is proportional to the geodesic curvature $k_g=1$ of the circle, with the normalization fixed by the Gauss--Bonnet theorem, $\tfrac1{4\pi}\big[\int_{\rm bulk}\sqrt{\hat g}\,\hat R+2\oint k_g\,\dd s\big]=\chi(\mathbb{RP}^2)=1$, in which the bulk term vanishes. Since $\varphi_*$ is constant on the circle and $g(1)=2$, we find
\begin{equation}
\frac1{8\pi}\int\sqrt{\hat g}\,\hat R\,\varphi_*
=\frac1{4\pi}\oint_{r=1}k_g\,\varphi_*\,\dd s
=\frac12\,\varphi_*(1)=-\log c-\log2 .
\end{equation}

When we add the counterterm $2\eta^2\log\epsilon$ to $-S_{\mathrm{cl}}^{(n)}=\eta\varphi_*(\epsilon)-(\text{kinetic}+\text{cosmological}+\text{curvature})$, the $\log\epsilon$ terms cancel, as do the two $\log2$ terms. The result is $-S_{\mathrm{cl}}^{(n)}=\gc\log c+\gc$, which is \eqref{eq:Scl-assemble}.

We can check this result with the Hellmann--Feynman theorem. Since the action is stationary at the saddle, the total $\eta$-derivative of $-S_{\mathrm{cl}}^{(n)}$ comes only from the explicit $\eta$-dependence of the source term and the counterterm, which gives $\varphi_*(\epsilon)+4\eta\log\epsilon=\varphi_*(0)=-2\log c$. This agrees with the derivative of \eqref{eq:Scl-assemble},
\begin{equation}
\frac{\dd(-S_{\mathrm{cl}})}{\dd\eta}=-2\frac{\dd}{\dd\gc}\big[\gc\log c+\gc\big]=-2\log c .
\end{equation}
Similarly, only the cosmological-constant term depends explicitly on $\log\Lambda$, so
\begin{equation}
\frac{\dd(-S_{\mathrm{cl}})}{\dd\log\Lambda}=\frac\gc2
=-\frac{\Lambda}{\pi}\int\dd^2z\,\ee^{\varphi_*}.
\end{equation}

\subsection{Saddle at fixed area}
\label{app:fa}

The fixed-area saddle has the form \eqref{eq:app-phi} with a real constant $c$, now determined by the area, $c^2=\pi/(2A\gc)$. The kinetic term does not depend on $c$, and the source and curvature terms depend on it only through $-2\log c$, so these three terms are given by the same expressions as in \eqref{eq:terms}. There is no cosmological-constant term at fixed area. With the counterterm added as before, we find $-S_{\mathrm{cl}}(A)=\gc\log c+\tfrac\gc2$, which is \eqref{eq:fa_assemble}. Substituting $\log c=-\tfrac12\log(2\gc A/\pi)$ gives \eqref{eq:fa_result}.

Alternatively, \eqref{eq:fa_result} follows from the Hellmann--Feynman theorem without evaluating the action on the solution. In the constrained action $\mathcal{S}^{(0)}[\varphi]+\tfrac{\Lambda_{\rm eff}}{\pi}\big(\int\ee^{\varphi}-A\big)$ the Lagrange multiplier plays the role of a cosmological constant, so that on shell the gradient of $\eta\varphi(0)-\mathcal{S}^{(0)}[\varphi]$ is $\tfrac{\Lambda_{\rm eff}}{\pi}\ee^{\varphi_*}$. The derivatives of the on-shell action therefore come only from its explicit dependence on $A$ and $\eta$,
\begin{equation}\label{eq:fa_hf}
\frac{\partial S_{\mathrm{cl}}}{\partial A}\Big|_\eta=-\frac{\Lambda_{\rm eff}}{\pi}=\frac{\gc}{2A},
\qquad
\frac{\partial S_{\mathrm{cl}}}{\partial \eta}\Big|_A=-\varphi_*(0)\big|_A=-\log\frac{2\gc A}{\pi}.
\end{equation}
The second equalities follow from \eqref{eq:fa_Leff} and \eqref{eq:fa_conetip}. The mixed derivatives agree, $\partial_\eta\partial_A S_{\mathrm{cl}}=\partial_A\partial_\eta S_{\mathrm{cl}}=-1/A$. Integrating \eqref{eq:fa_hf} reproduces \eqref{eq:fa_result} up to an additive constant.

\section{The Gelfand--Yaglom theorem and the crosscap boundary conditions}
\label{app:gy}

In this appendix we state the Gelfand--Yaglom theorem in the form used in section~\ref{sec:detD}. After separation of variables, each radial operator $D_m$ acts on functions on $0\le r\le1$ that behave as $r^{|m|}$ at the singular endpoint $r=0$ and satisfy the crosscap condition \eqref{eq:bc} at $r=1$, which is Neumann for even $m$ and Dirichlet for odd $m$.

For a Sturm--Liouville operator on an interval, the Gelfand--Yaglom theorem expresses the determinant in terms of the boundary value at $r=1$ of the solution $y_m$ that is regular at $r=0$, up to a constant that is common to $D_m$ and to the free operator $D_m^0=-\partial_r^2-r^{-1}\partial_r+m^2/r^2$. The boundary value that appears depends on the condition at $r=1$,
\begin{equation}\label{eq:gy-data}
\det D_m\ \propto\ y_m(1)\ \ (\text{Dirichlet}),
\qquad
\det D_m\ \propto\ y_m'(1)\ \ (\text{Neumann}).
\end{equation}
In the $m=0$ sector we use \eqref{eq:gy-data} as it stands and leave the constant undetermined. For $m\neq0$ the constant cancels in the ratio to the determinant of the free operator with the same boundary condition,
\begin{equation}\label{eq:gy-ratio}
\frac{\det D_m}{\det D_m^0}=\frac{y_m(1)}{y_m^{0}(1)}\ \ (\text{Dirichlet}),
\qquad
\frac{\det D_m}{\det D_m^0}=\frac{y_m'(1)}{y_m^{0\prime}(1)}\ \ (\text{Neumann}),
\end{equation}
where both regular solutions are normalized to the same leading behavior $y_m\sim r^{|m|}$ as $r\to0$.

The regular solution of the free operator is $y_m^0=r^{|m|}$, with $y_m^0(1)=1$ and $y_m^{0\prime}(1)=|m|$. The Neumann boundary values of the two regular solutions grow like $|m|$ at large $|m|$, but this growth cancels in the ratio. Substituting the regular solution \eqref{eq:reg-sol} of $D_m\psi=0$, normalized to unit leading coefficient, into \eqref{eq:gy-ratio} gives the ratios \eqref{eq:per-mode}. Both behave as $1-\gc/|m|$ at large $|m|$, so the sums in \eqref{eq:sums} diverge only logarithmically.

\section{An independent evaluation of the fluctuation determinant}
\label{app:renorm}

In section~\ref{sec:oneloop} we obtained the factors $\Gamma(2\eta)$ and $\sin\pi\eta$ and the coefficient of $\log\gc$ without reference to a normalization scale, but the $\eta$-linear term $\gc\log2$ in \eqref{eq:detD} depends on the choice of the reference operator $D_m^0$ and on the regularization of the mode sums. In this appendix we evaluate $\Det D$ in a second way, with the variational formula of Menotti and Tonni~\cite{MenottiTonniI,MenottiTonniII}. This formula gives the $\eta$-derivative of $\log\Det D$ in terms of the diagonal of the Green function, in a renormalization scheme defined by the subtraction of its short-distance singularity. The part of the result that does not depend on the renormalization scheme agrees with that of section~\ref{sec:oneloop}. In appendix~\ref{sec:mt-crosscheck} we fix the $\eta$-linear term, which depends on the renormalization scheme, by matching the normalization of the cosmological constant to that of the exact formula.

\subsection{The variational formula}
\label{sec:mt-formula}

By the Jacobi formula, $\partial_\eta\log\Det D=\Tr(D^{-1}\partial_\eta D)$. At fixed $\Lambda$ the operator $D=-\Delta_\delta+8\Lambda\ee^{\varphi_*}$ depends on $\eta$ only through its potential, so $\partial_\eta D=8\Lambda\,\partial_\eta\ee^{\varphi_*}$ is a multiplication operator. For a multiplication operator $f$ the trace is $\Tr(D^{-1}f)=\tfrac1{2\pi}\int G(z,z)\,f\,\dd^2z$, where $G$ is the Green function normalized by $D_z G(z,w)=2\pi\delta^2(z-w)$. Replacing the divergent $G(z,z)$ by its regularized value \eqref{eq:mt-reg} defined below, we obtain
\begin{equation}\label{eq:mt-var}
\partial_\eta\log(\Det D)^{-1/2}
=-\frac{2\Lambda}{\pi}\int G(z,z)_{\rm reg}\,\partial_\eta\ee^{\varphi_*}\,\dd^2z.
\end{equation}
With this normalization $G(z,w)=-\log|z-w|+O(1)$ as $w\to z$, so we define the regularized diagonal value by
\begin{equation}\label{eq:mt-reg}
G(z,z)_{\rm reg}=\lim_{w\to z}\big[\,G(z,w)+\log|z-w|\,\big].
\end{equation}

Since \eqref{eq:mt-reg} subtracts the logarithm of the flat distance $|z-w|$, ultraviolet divergences are regularized with respect to the flat metric $\delta$, as in the one-loop computation on the pseudosphere~\cite{ZZpseudosphere}. Subtracting the logarithm of the geodesic distance in the metric $\ee^{\varphi_*}\delta$ instead would shift $G(z,z)_{\rm reg}$ by a local term proportional to $\varphi_*$. Subtracting the logarithm of the flat distance is consistent with the bulk DOZZ normalization~\cite{DornOtto,ZZbootstrap,TeschnerReview}. We return to this point in appendix~\ref{sec:mt-crosscheck}.

\subsection{The Green function on the sphere}
\label{sec:mt-sphere}

We first construct the components $G_m$ with $m\neq0$ on the sphere and pass to $\mathbb{RP}^2$ in appendix~\ref{sec:mt-rp2}. By rotational symmetry $G(z,w)=\sum_m G_m(r,r_0)\ee^{im(\theta-\theta_0)}$. Each radial Green function $G_m$ is constructed from two solutions of the P\"oschl--Teller equation of section~\ref{sec:detD}: the solution $f_+$ regular at $r=0$, which is the $f_m$ of \eqref{eq:reg-sol}, and the solution $f_-$ regular at $r=\infty$. With $r_<=\min(r,r_0)$, $r_>=\max(r,r_0)$, and the Wronskian $r\,W[f_+,f_-]=-2|m|(1-m^2/\gc^2)$, we find
\begin{equation}\label{eq:mt-green}
G_m(r,r_0)=\frac{f_+(r_<)\,f_-(r_>)}{2|m|(1-m^2/\gc^2)},
\qquad
f_\pm=\Big(\tanh(\gc\log r)\mp\frac{|m|}{\gc}\Big)\,r^{\pm|m|}.
\end{equation}
On the diagonal, with $\tau=\log r$, this becomes
\begin{equation}\label{eq:mt-diag}
G_m(r,r)=\frac{1}{2|m|}-\frac{\sech^2(\gc\tau)}{2|m|(1-m^2/\gc^2)}.
\end{equation}
The first term is the diagonal value of the free Green function, obtained by setting the potential of $D_m$ to zero. Summed over $m\neq0$, the free Green function behaves as $-\log|z-w|+\log r$ near the diagonal. The subtraction \eqref{eq:mt-reg} removes $-\log|z-w|$, while $\log r$ integrates to zero in \eqref{eq:mt-var}, both on the sphere and over $r\le1$. We sum the second term of \eqref{eq:mt-diag} with the digamma-function identity
\begin{equation}\label{eq:mt-digamma}
\sum_{m\ge1}\frac{1}{m(1-m^2/\gc^2)}
=\frac12\big(\psi(1-\gc)+\psi(1+\gc)\big)+\gamma_E
\end{equation}
and obtain the regularized diagonal value on the sphere, with the $m=0$ mode excluded and the term $\log r$ dropped,
\begin{equation}\label{eq:mt-Bsphere}
G(z,z)^{\mathbb{S}^2}_{\rm reg}\big|_{m\ne0}=-\sech^2(\gc\tau)\,B,
\qquad
B\equiv\frac12\big(\psi(1-\gc)+\psi(1+\gc)\big)+\gamma_E .
\end{equation}

To evaluate \eqref{eq:mt-var} with \eqref{eq:mt-Bsphere}, we use $\partial_\eta\ee^{\varphi_*}=\gc\,\Lambda^{-1}r^{-2}\,\sech^2(\gc\tau)\,[1-\gc\tau\tanh(\gc\tau)]$ and $\dd^2z/r^2=\dd\tau\,\dd\theta$. The angular integral gives $2\pi$, and the radial integral is
\begin{equation}
\int_{-\infty}^{\infty}\sech^4(\gc\tau)\,\big[1-\gc\tau\tanh(\gc\tau)\big]\,\dd\tau
=\frac{1}{\gc}.
\end{equation}
Since this factor $1/\gc$ cancels the explicit $\gc$ in $\partial_\eta\ee^{\varphi_*}$, we find
\begin{equation}\label{eq:mt-sphere}
\partial_\eta\log(\Det D)^{-1/2}\big|_{\mathbb{S}^2,\,m\ne0}=4B=2\psi(2\eta)+2\psi(2-2\eta)+4\gamma_E .
\end{equation}
The term $4\gamma_E$, which comes from the constant $\gamma_E$ in $B$, integrates to an $\eta$-linear term.

\subsection{Projection to the real projective plane}
\label{sec:mt-rp2}

Since $\sigma:z\mapsto-1/\bar z$ exchanges the interior and the exterior of the unit disk, the Green function on $\mathbb{RP}^2$ follows from the one on the sphere by the method of images, $G_{\mathbb{RP}^2}(z,w)=G_{\mathbb{S}^2}(z,w)+G_{\mathbb{S}^2}(z,\sigma w)$. For $m\neq0$ the regularized diagonal value in \eqref{eq:mt-var} is therefore
\begin{equation}\label{eq:mt-image}
G_{\mathbb{RP}^2}(z,z)_{\rm reg}=\underbrace{G_{\mathbb{S}^2}(z,z)_{\rm reg}}_{\text{direct}}
+\underbrace{G_{\mathbb{S}^2}(z,\sigma z)}_{\text{antipodal}} ,
\end{equation}
where the antipodal term has no short-distance singularity since $\sigma z\ne z$.

The $m=0$ sector must be treated separately. On the sphere the dilatation mode $\tanh(\gc\tau)$ solves $D_0u=0$ and is regular at both poles, so $D_0$ has a zero mode and no Green function. On $\mathbb{RP}^2$ the same solution violates the Neumann condition at $r=1$, so $D_0$ has no zero mode and does have a Green function $G_0$. We therefore evaluate three contributions, each integrated over the fundamental domain $r\le1$: the direct term, the antipodal term, and the $m=0$ term computed on $\mathbb{RP}^2$. The integrand of the direct term in \eqref{eq:mt-var} is even in $\tau$, so its integral over $r\le1$ is half of \eqref{eq:mt-sphere}.

The antipodal term is $G_{\mathbb{S}^2}(z,\sigma z)=2\sum_{m\ge1}(-1)^m G_m(r,1/r)$, where the sign $(-1)^m$ comes from the shift $\theta\to\theta+\pi$. With $f_+(r)\,f_-(1/r)=-(\tanh\gc\tau-|m|/\gc)^2\,\ee^{2|m|\tau}$ for $r\le1$, the contribution of the antipodal term to \eqref{eq:mt-var} involves the radial integral
\begin{equation}\label{eq:mt-Jm}
J_m=\int_{-\infty}^0\Big(\tanh\gc\tau-\frac{|m|}\gc\Big)^2\sech^2\gc\tau\,\big[1-\gc\tau\tanh\gc\tau\big]\,\ee^{2|m|\tau}\,\dd\tau=\frac{m}{2\gc^2}.
\end{equation}
The factor $m$ in $J_m$ cancels the $1/m$ in \eqref{eq:mt-green}, leaving an alternating sum,
\begin{equation}\label{eq:mt-anti}
\partial_\eta\log(\Det D)^{-1/2}\big|_{\rm anti}
=2\gc\sum_{m\ge1}\frac{(-1)^m}{\gc^2-m^2}
=\frac{\pi}{\sin\pi\gc}-\frac1{\gc},
\end{equation}
where we used the Mittag--Leffler expansion of $\pi/\sin\pi\gc$ and separated its pole term $1/\gc$. Combined with the direct term, this alternating sum produces the factor $\sin\pi\eta$.

In the $m=0$ sector we construct $G_0$ from $u_<=\tanh(\gc\tau)$, which is regular at $r=0$, and $u_>=1-\gc\tau\tanh(\gc\tau)$, which satisfies the Neumann condition at $r=1$. Their Wronskian is $r\,W[u_<,u_>]=-\gc$, so $G_0(r,r)=\tanh(\gc\tau)[1-\gc\tau\tanh(\gc\tau)]/\gc$. When we substitute this into \eqref{eq:mt-var}, the explicit $\gc$ cancels again, and we find
\begin{equation}\label{eq:mt-m0}
\partial_\eta\log(\Det D)^{-1/2}\big|_{m=0}
=-\frac{4}{\gc}\int_{-\infty}^{0}\tanh x\,\sech^2x\,(1-x\tanh x)^2\,\dd x
=\frac{1}{1-2\eta},
\end{equation}
where the integral equals $-\tfrac14$.

In the direct term we write $\psi(2-2\eta)=\psi(1-2\eta)+1/(1-2\eta)$, whose last term cancels the $-1/\gc$ of the antipodal term. The total, including the $m=0$ term \eqref{eq:mt-m0}, is then
\begin{equation}\label{eq:mt-total}
\partial_\eta\log(\Det D_{\mathbb{RP}^2})^{-1/2}
=\psi(2\eta)+\psi(1-2\eta)+\frac{\pi}{\sin2\pi\eta}+2\gamma_E+\frac{1}{1-2\eta}.
\end{equation}
Combining the first three terms into $2\psi(2\eta)+\pi\cot\pi\eta$ with the reflection formula and integrating with respect to $\eta$, we obtain
\begin{equation}\label{eq:mt-result}
-\tfrac12\log\Det D_{\mathbb{RP}^2}
=\log\Gamma(2\eta)+\log\sin\pi\eta-\tfrac12\log\gc+2\gamma_E\eta+\text{const}.
\end{equation}
The coefficient $2\gamma_E$ of the $\eta$-linear term is half of the sphere value $4\gamma_E$ in \eqref{eq:mt-sphere} because the direct term is integrated only over the fundamental domain, while the antipodal and $m=0$ terms contain no $\gamma_E$.

\subsection{Cross-check, renormalization scheme dependence, and the cosmological constant}
\label{sec:mt-crosscheck}

In exponentiated form, the result \eqref{eq:detD} of section~\ref{sec:detD} and the result \eqref{eq:mt-result} of this appendix read
\begin{equation}\label{eq:mt-compare}
2^{\gc}\,\Gamma(2\eta)\sin\pi\eta\,\gc^{-1/2}\ \ (\text{Gelfand--Yaglom}),
\qquad
\ee^{2\gamma_E\eta}\,\Gamma(2\eta)\sin\pi\eta\,\gc^{-1/2}\ \ (\text{this appendix}).
\end{equation}
They differ only in the $\eta$-linear factor, $2^{\gc}$ as opposed to $\ee^{2\gamma_E\eta}$, as required by locality.

In two dimensions, if the flat metric is kept as the reference, a change of renormalization scheme shifts $\log\Det D$ by a local counterterm $\int\sqrt{\hat g}\,(c_1\hat R+c_2 V)$, where $V=8\Lambda\ee^{\varphi_*}$ is the potential of $D$. This counterterm has the same form as the integrated Seeley--DeWitt coefficient $a_1$. Here $\int\sqrt{\hat g}\,\hat R=4\pi\chi(\mathbb{RP}^2)$ does not depend on $\eta$, while $\int\sqrt{\hat g}\,V=8\Lambda\int\ee^{\varphi_*}=-4\pi\gc$ is linear in $\eta$. A change of renormalization scheme can therefore shift $\log\Det D$ only by a term linear in $\eta$. In particular, the transcendental factors $\Gamma(2\eta)$ and $\sin\pi\eta$ do not depend on the renormalization scheme. In $-\tfrac12\log\Det D$ the $\eta$-linear term is $\gc\log2$ for the free reference operator and the zeta regularization of section~\ref{sec:detD}, and $2\gamma_E\eta$ for the subtraction \eqref{eq:mt-reg}.

The value $2\gamma_E\eta$ obtained with the subtraction \eqref{eq:mt-reg} matches the normalization of the cosmological constant in the exact formula. The formula \eqref{eq:U} contains the multiplicative renormalization $\mu\to\mu\,\gamma(b^2)$, with $\gamma(b^2)=b^{-2}-2\gamma_E+O(b^2)$. In $\log(\pi\mu\,\gamma(b^2))^{\nu+1/2}$ the term $-2\gamma_E$ contributes $\nu\cdot(-2\gamma_E b^2)=-\gc\gamma_E$ at $O(b^0)$, whose $\eta$-dependent part $2\gamma_E\eta$ agrees with \eqref{eq:mt-result}. The terms with $\log b$ in the same factor cancel against those from the Stirling expansion of $\Gamma(\gc/b^2+2)$ in \eqref{eq:Gf} and from the prefactor $2/b$ in \eqref{eq:U}.

\bibliographystyle{JHEP}
\bibliography{main}

\end{document}